\documentstyle[aps,epsfig,preprint]{revtex}

\begin{document}

\vspace{1.0cm}
\begin{center}
  {\Huge \bf Energy Dependence of}  
\end{center}
\begin{center}
  {\Huge \bf  Pion and Kaon Production} 
\end{center}
\begin{center}
  {\Huge \bf  in Central Pb+Pb Collisions} 
\end{center}

\vspace{0.5cm}

\begin{center}
  {\Large \bf The NA49 Collaboration}
\end{center}


\begin{abstract}
\noindent
Measurements of charged pion and kaon production
in central Pb+Pb collisions at 40, 80  and 158 $A$GeV 
are presented.
These are compared with  data at lower and higher
energies as well as with results from $p+p$ interactions.
The mean pion multiplicity per wounded nucleon increases approximately
linearly with $s_{NN}^{1/4}$ with a change of slope starting
in the region 15--40 $A$GeV.
The change from pion suppression  
with respect to $p+p$ interactions, 
as observed at low collision energies, to pion enhancement
at high energies  occurs at about 40 $A$GeV.
A non--monotonic energy dependence of the 
ratio of $K^+$ to $\pi^+$ yields is observed, with 
a maximum close to 40 $A$GeV and an indication of a nearly
constant value at higher energies. 
The measured dependences may be related to an increase of the entropy
production and a decrease of the strangeness to entropy ratio
in central Pb+Pb collisions in the low SPS energy range, which is consistent
with the hypothesis that  a transient state of deconfined
matter is created above these energies.
Other interpretations of the data are also discussed.

\end{abstract}

\newpage

\begin{center}
  {\bf The NA49 Collaboration}
\end{center}
\vspace{0.5cm}
\noindent
S.V.~Afanasiev$^{9}$,T.~Anticic$^{20}$, D.~Barna$^{5}$,
J.~Bartke$^{7}$, R.A.~Barton$^{3}$, M.~Behler$^{15}$, 
L.~Betev$^{10}$, H.~Bia{\l}\-kowska$^{17}$, A.~Billmeier$^{10}$,
C.~Blume$^{8}$, C.O.~Blyth$^{3}$, B.~Boimska$^{17}$, M.~Botje$^{1}$,
J.~Bracinik$^{4}$, R.~Bramm$^{10}$, R.~Brun$^{11}$,
P.~Bun\v{c}i\'{c}$^{10,11}$, V.~Cerny$^{4}$,
J.G.~Cramer$^{19}$, P.~Csat\'{o}$^{5}$, P.~Dinkelaker$^{10}$,
V.~Eckardt$^{16}$, P.~Filip$^{16}$,
Z.~Fodor$^{5}$, P.~Foka$^{8}$, P.~Freund$^{16}$,
V.~Friese$^{15}$, J.~G\'{a}l$^{5}$,
M.~Ga\'zdzicki$^{10}$, G.~Georgopoulos$^{2}$, E.~G{\l}adysz$^{7}$, 
S.~Hegyi$^{5}$, C.~H\"{o}hne$^{15}$, G.~Igo$^{14}$,
P.G.~Jones$^{3}$, K.~Kadija$^{11,20}$, A.~Karev$^{16}$,
V.I.~Kolesnikov$^{9}$, T.~Kollegger$^{10}$, M.~Kowalski$^{7}$, 
I.~Kraus$^{8}$, M.~Kreps$^{4}$, M.~van~Leeuwen$^{1}$, 
P.~L\'{e}vai$^{5}$, A.I.~Malakhov$^{9}$, S.~Margetis$^{13}$,
C.~Markert$^{8}$, B.W.~Mayes$^{12}$, G.L.~Melkumov$^{9}$,
A.~Mischke$^{8}$, J.~Moln\'{a}r$^{5}$, J.M.~Nelson$^{3}$,
G.~P\'{a}lla$^{5}$, A.D.~Panagiotou$^{2}$,
K.~Perl$^{18}$, A.~Petridis$^{2}$, M.~Pikna$^{4}$, L.~Pinsky$^{12}$,
F.~P\"{u}hlhofer$^{15}$,
J.G.~Reid$^{19}$, R.~Renfordt$^{10}$, W.~Retyk$^{18}$,
C.~Roland$^{6}$, G.~Roland$^{6}$, A.~Rybicki$^{7}$, T.~Sammer$^{16}$,
A.~Sandoval$^{8}$, H.~Sann$^{8}$, N.~Schmitz$^{16}$, P.~Seyboth$^{16}$,
F.~Sikl\'{e}r$^{5}$, B.~Sitar$^{4}$, E.~Skrzypczak$^{18}$,
G.T.A.~Squier$^{3}$, R.~Stock$^{10}$, H.~Str\"{o}bele$^{10}$, T.~Susa$^{20}$,
I.~Szentp\'{e}tery$^{5}$, J.~Sziklai$^{5}$,
T.A.~Trainor$^{19}$, D.~Varga$^{5}$, M.~Vassiliou$^{2}$,
G.I.~Veres$^{5}$, G.~Vesztergombi$^{5}$,
D.~Vrani\'{c}$^{8}$, A.~Wetzler$^{10}$, C.~Whitten$^{14}$,
I.K.~Yoo$^{15}$, J.~Zaranek$^{10}$, J.~Zim\'{a}nyi$^{5}$

\vspace{0.5cm}
\noindent
$^{1}$NIKHEF, Amsterdam, Netherlands. \\
$^{2}$Department of Physics, University of Athens, Athens, Greece.\\
$^{3}$Birmingham University, Birmingham, England.\\
$^{4}$Comenius University, Bratislava, Slovakia.\\
$^{5}$KFKI Research Institute for Particle and Nuclear Physics, Budapest, Hungary.\\
$^{6}$MIT, Cambridge, USA.\\
$^{7}$Institute of Nuclear Physics, Cracow, Poland.\\
$^{8}$Gesellschaft f\"{u}r Schwerionenforschung (GSI), Darmstadt, Germany.\\
$^{9}$Joint Institute for Nuclear Research, Dubna, Russia.\\
$^{10}$Fachbereich Physik der Universit\"{a}t, Frankfurt, Germany.\\
$^{11}$CERN, Geneva, Switzerland.\\
$^{12}$University of Houston, Houston, TX, USA.\\
$^{13}$Kent State University, Kent, OH, USA.\\
$^{14}$University of California at Los Angeles, Los Angeles, USA.\\
$^{15}$Fachbereich Physik der Universit\"{a}t, Marburg, Germany.\\
$^{16}$Max-Planck-Institut f\"{u}r Physik, Munich, Germany.\\
$^{17}$Institute for Nuclear Studies, Warsaw, Poland.\\
$^{18}$Institute for Experimental Physics, University of Warsaw, Warsaw, Poland.\\
$^{19}$Nuclear Physics Laboratory, University of Washington, Seattle, WA, USA.\\
$^{20}$Rudjer Boskovic Institute, Zagreb, Croatia.\\

\vspace{0.5cm}
\newpage
\section{Introduction}
The primary purpose of the heavy ion programme at the CERN SPS is 
the search for a transient deconfined state
of strongly interacting matter during the early stage of 
nucleus--nucleus collisions \cite{qm99}. When a sufficiently high
initial energy density is reached, the formation of such a state
of quasi free quarks and gluons, the quark gluon plasma (QGP),
is expected. 
A key problem is the identification of 
experimental signatures of QGP creation \cite{qgp}. 
Numerous proposals were discussed in the past \cite{Ra:82},
but the significance of these signals has come under
renewed scrutiny. 
A possible, promising strategy is a study of the energy
dependence of pion and strangeness yields.
It
was suggested \cite{GaRo,Ga:95,GaGo} that the transition may lead
to anomalies in this dependence:  
a steepening of the increase of the pion yield and 
a non monotonic behaviour of the strangeness to pion ratio.
The space--time evolution
of the created fireball \cite{Hu:95} and  the event--by--event
fluctuations \cite{St:99} may also be sensitive to
crossing the transition region. 

First experimental results from Pb+Pb (Au+Au) collisions at
top SPS (158 $A$GeV) and AGS (11 $A$GeV) energies
have suggested \cite{GaRo}
that anomalies in pion and strangeness production  may be located
between these energies. 
The study of this hypothesis is the motivation for a dedicated  
energy scan at the CERN SPS \cite{na49_add1}. 
Within this ongoing project NA49 has recorded central 
Pb+Pb collisions at 40 and 80 $A$GeV during the heavy ion runs in
1999 and 2000, respectively. The data at the top SPS energy 
(158 $A$GeV) were taken in previous SPS runs.
In this paper we report  final results on the energy
dependence of charged pion and kaon production.  
A preliminary analysis was presented
in Refs. \cite{Bl:01,Ko:01}.
Pseudorapidity spectra of charged particles
produced in Pb+Pb collisions at 40 and 158 $A$GeV
were recently published by the NA50 experiment
\cite{na50}.    

The energy scan programme at the CERN SPS will be completed in 2002
by taking data at 20 and 30 $A$GeV.

\section{Experimental Set--up}
\begin{figure}[p]

\vspace{2cm}

\begin{center}
\epsfig{file= 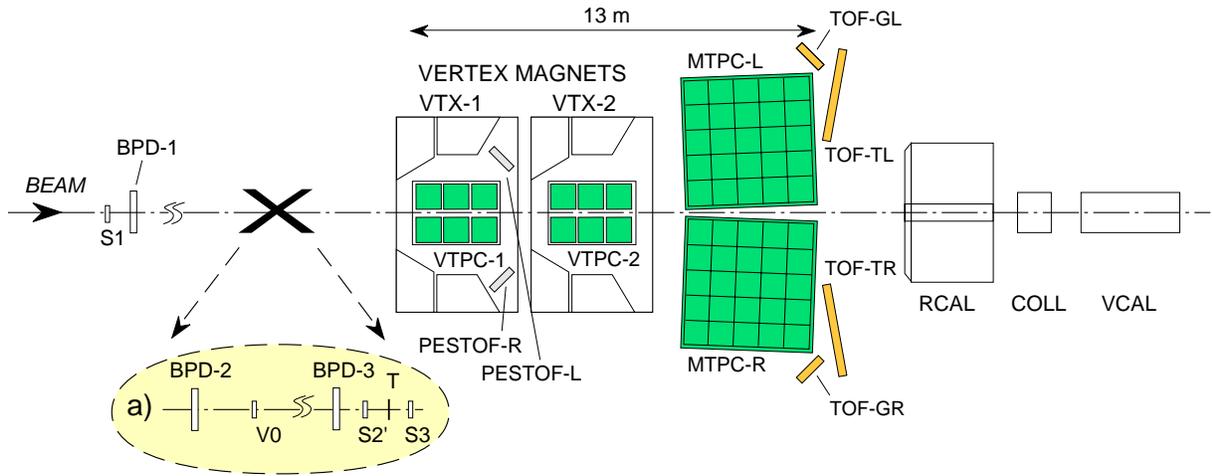,width=16cm}
\end{center}

\vspace{0.5cm}

\caption{The experimental set--up of the NA49 experiment 
\protect\cite{na49_nim}.
}
\label{setup}
\end{figure}

The NA49 experimental set--up  \cite{na49_nim} is shown in
Fig. \ref{setup}.
It consists of four large volume 
Time Projection Chambers (TPCs).
Two of these, the Vertex TPCs (VTPC-1 and VTPC-2), are placed in the
magnetic field of two super-conducting dipole  magnets.
This allows separation of positively and negatively charged
tracks and a precise measurement of the particle momenta $p$ 
with a resolution of
$\sigma(p)/p^2 \cong (0.3-7)\cdot10^{-4}$ (GeV/c)$^{-1}$.  
The other two TPCs (MTPC-L and MTPC-R), positioned 
downstream of the magnets were optimised for high
precision measurement of the ionization energy loss $dE/dx$ (relative 
resolution of about 4\%) which provides 
a means of determining the particle mass. 
The particle identification capability of the MTPCs is
augmented by two Time of Flight (TOF)
detector arrays with a resolution
$\sigma_{tof} \cong $ 60 ps. 

\begin{figure}[p]

\hspace{3cm}

\begin{center}
\epsfig{file=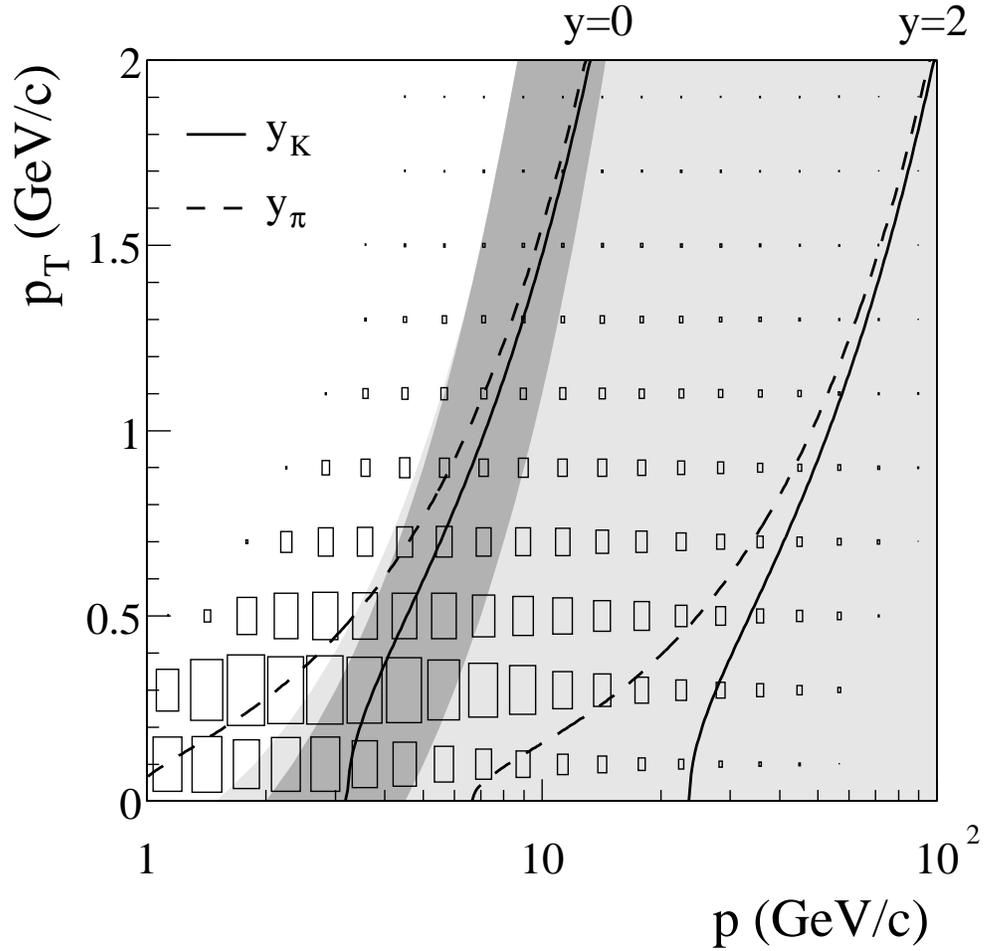,width=14cm}
\end{center}

\vspace{1.5cm}

\caption{
Distribution of accepted unidentified particles versus the total
momentum $p$ and the transverse momentum $p_T$ at 80~$A$GeV beam
energy. The light (dark) shaded area indicates the acceptance of
the MTPCs (TOF detectors).
Also shown are two isolines of constant
rapidity $y$ for kaons (full curves) and pions (dashed curves).
  }
\label{acceptance}
\end{figure}

The acceptance of the NA49 detector is illustrated in  Fig.~\ref{acceptance}
where we show the distribution of reconstructed unidentified
particles
versus the total momentum $p$ and transverse momentum
$p_T$ for the 80~$A$GeV data. 
The light (dark) shaded region shows
the coverage of the MTPCs (TOF) detectors. The resolution of the
$dE/dx$ 
measurement allows hadron identification in the MTPCs for
$p > 4$~GeV. At each incident energy
the TOF acceptance for kaons was kept at midrapidity by lowering
the nominal magnetic field ($B$(VTX--1,~2) $\approx (1.5,~1.1)$~T at
158~$A$GeV) in proportion to the beam energy. Data were taken for
both field polarities.

The target, a thin lead foil (224 mg/cm$^2$ $\cong$1\%
of the interaction length), was positioned about
80 cm upstream from VTPC-1.

Central collisions were selected by a trigger using information
from a downstream calorimeter (VCAL), which measured the energy 
of the projectile spectator nucleons.
The geometrical acceptance of the VCAL calorimeter
was adjusted for each energy in order to cover the projectile spectator
region 
by a proper setting of
a collimator (COLL)~\cite{na49_nim,veto}. 

\section{Analysis}
Raw $K^+$ and  $K^-$ yields were extracted from  fits of the 
distributions of $dE/dx$ and $tof$ (where available) in narrow 
bins of momentum and transverse momentum. 
The spectra at midrapidity are obtained using the combined $dE/dx$ and
$tof$ information ($tof+dE/dx$ analysis). 
The resulting distributions were 
corrected for geometrical acceptance, losses due to in--flight decays and
reconstruction efficiency. 
The first two corrections are calculated using the  
detector simulation package GEANT \cite{geant}.
The procedure used for the efficiency calculation is discussed below,
in the context of the pion analysis.
 
For pions at midrapidity the acceptance of the $dE/dx$ 
and $tof+dE/dx$ methods is limited
to the high $p_T$ region  (see Fig.~\ref{acceptance}).
To obtain the $\pi^-$ spectra yields of all negatively charged particles 
were determined as a function of rapidity  
(calculated assuming the $\pi$--mass) and $p_T$.
The contamination of  $K^-$, $\overline{p}$
and $e^-$ from the interaction vertex
as well as non--vertex hadrons originating from strange
particle decays and secondary interactions was subtracted
using two different methods. 

In the first method each track measured in the TPCs was extrapolated back to the
target plane. The distance between the track and the interaction vertex was
calculated in this plane (track impact parameter). 
The impact parameter distributions were used to establish cuts
for  selection of tracks from the interaction vertex and to
estimate the contribution of remaining non--vertex tracks.
A correction for $K^-$ contamination was calculated using
parametrised $K^-$ spectra. 

In the second method all tracks fitted to the interaction vertex
were accepted. 
The necessary corrections were calculated based on a Venus
\cite{venus} simulation of central Pb+Pb collisions. 
The Venus events were processed
by GEANT \cite{geant}
and the NA49 software which simulates the TPC response and produces
files in raw data format. These events were reconstructed and
the reconstructed tracks were matched to the Venus input.
The obtained correction was scaled by a factor which matches 
the simulated Venus and the measured 
hadron yields.
The corrections determined using  both methods are compatible.
The Venus based correction, amounting to 20--25\%,
is used for the data presented in this
paper.

The $\pi^+$ spectra were not analysed, because the positive particles
have a large and uncertain 
contribution of protons.

The resulting $K^{\pm}$ and $\pi^-$ spectra were corrected for geometrical 
acceptance,  and losses due to inefficiencies of the
tracking algorithms and quality cuts. These losses were determined using 
the `embedding' method. Events containing a few tracks were generated
and processed by the simulation software. The resulting raw data were
embedded into real events.
The combined raw data were reconstructed 
and the input tracks were matched with the reconstructed ones. The
calculated losses are about~5\%.

In order to reduce the systematic errors, the analysis has been
restricted to regions of phase space where  the background and
efficiency corrections are small and approximately
uniform.  The systematic errors were estimated to be below 10\%.  This
estimate is based on the comparison of results obtained 
using different detectors (TOF, TPCs), and varying cuts and correction
strategies (see above). Additionally, data taken at the two magnetic
field polarities were analysed and the results were found to agree
within 5\%.
Note that the same experimental procedure was used
to obtain results at all three energies. 
Thus to a large extent the systematic uncertainties
are common for the NA49 measurements.

The average number of wounded nucleons \cite{bialas}
$\langle N_W \rangle$ 
(the notation $\langle \dots \rangle$ will be used to denote the mean 
multiplicity in full phase space throughout the paper) 
as given in  Table~I was not directly measured but  was
calculated using the Fritiof model \cite{Fr}.
In an unbiased sample of generated inelastic interactions
a subsample of central events was selected by applying a cut
in the number of projectile spectators.
This cut  
was adjusted such that the selected fraction was
equal to the 
fraction of all inelastic interactions accepted by the central
trigger in the experiment.
The value of $\langle N_W \rangle$ was then 
calculated for
the Fritiof Pb+Pb collisions 
selected in this way.
Finally it was verified that the $\langle N_W \rangle$ value
for central Pb+Pb collisions at 158 $A$GeV
agrees (within several per cent) 
with the total number of net baryons determined  from the participant
domain (in the rapidity interval $-2.6 < y < 2.6$) 
for these collisions \cite{PRL}. 

Note that in this paper $y$ denotes the rapidity of
a particle in the collision center--of--mass system.

Table I summarises the parameters characterising 
the data samples used in this analysis.

\section{Results}
\begin{figure}[p]

\hspace{3cm}

\begin{center}
\epsfig{file= 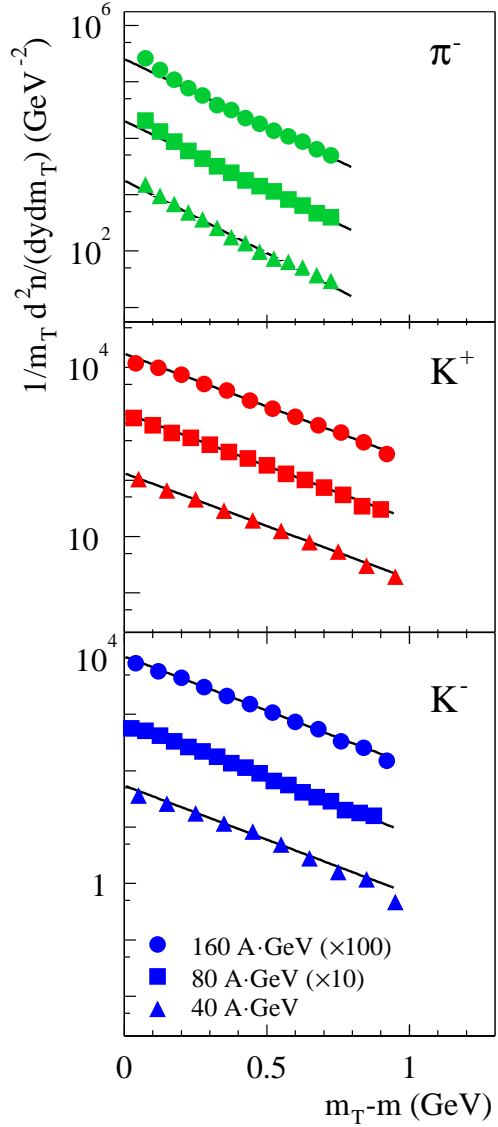,width=7cm}
\end{center}

\vspace{0.5cm}

\caption{Transverse mass spectra of
$\pi^-$, $K^+$ and $K^-$ mesons 
produced at midrapidity ($|y| < 0.1$ for kaons in the $tof+dE/dx$ analysis, and
$0 < y < 0.2$ for pions)
in
central Pb+Pb collisions at 40 (triangles), 80 (squares) 
and 158 (circles)
$A$GeV.
The values for 80 $A$GeV and 158 $A$GeV
are scaled by  factors of 10 and 100, respectively.
The lines are exponential fits to the spectra
(see text, Eq. 1) in the interval
0.2 GeV $< m_T -m <$ 0.7  GeV.
Statistical errors are smaller than the symbol size.
The systematic errors are $\pm$5\% in the region used for the fit
and reach $\pm$10\% at the edges of the acceptance.
}
\label{dndmt}
\end{figure}

Spectra of transverse mass  $m_T = \sqrt{p_T^2 + m^2}$
($m$ is the rest mass of the particle)
for $K^+$, $K^-$ ($tof+dE/dx$ analysis)
and $\pi^-$ mesons produced near midrapidity 
($|y| < 0.1$ for kaons and $0 < y < 0.2$ for pions) 
in central Pb+Pb collisions at 40, 80 and 158 $A$GeV are shown 
in Fig. \ref{dndmt}. The full lines indicate a fit of the
function  
\begin{equation}
\frac {dn} {m_T dm_T dy} = C \cdot \exp  \left( - \frac {m_T} {T} \right) 
\label{mt}
\end{equation}
to the data in the range 0.2 GeV $< m_T -m <$ 0.7  GeV. 
The values obtained for the inverse slope parameter $T$ 
are presented in Table II.
The $T$ parameter is smaller for pions than for kaons.
A weak increase of $T$ with increasing energy is suggested
by the pion data, whereas no significant change is seen for kaons.

\begin{figure}[p]

\hspace{2cm}
\begin{center}
\epsfig{file=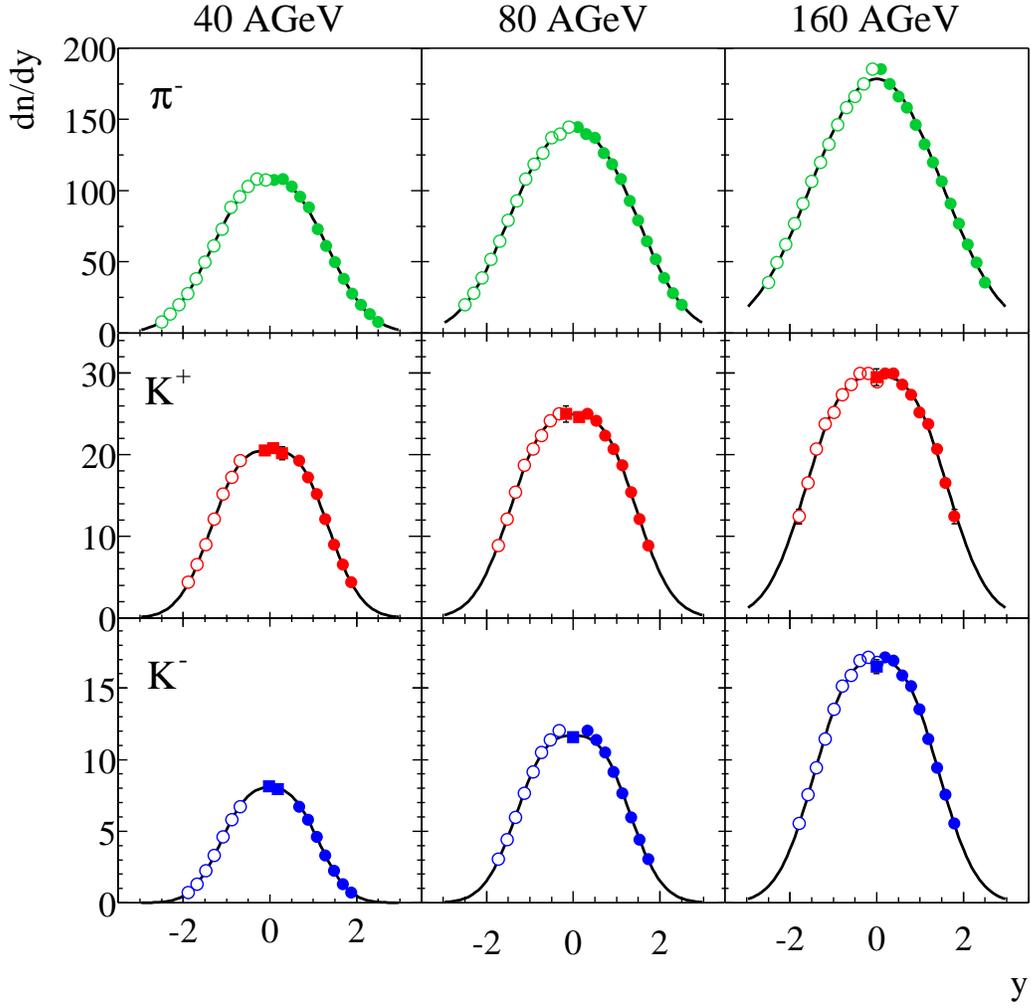,width=14cm}
\end{center}

\vspace{1.0cm}
\caption{Rapidity distributions of
$\pi^-$, $K^+$ and $K^-$ mesons produced in
central Pb+Pb collisions at 40, 
80 and 158 
$A$GeV.
For kaons
squares and circles indicate the results of $tof+dE/dx$  and
$dE/dx$ only analyses, respectively.
The closed symbols indicate measured points, open points are
reflected with respect to midrapidity. The lines indicate 
two--Gauss fits to the spectra (see Eq. 2).
The plotted errors,
which are mostly smaller than the symbol size, are statistical only, 
the systematic errors
are $\pm$5\%.
}
\label{dndy}
\end{figure}

The rapidity distributions $dn/dy$ plotted in Fig. \ref{dndy} were
obtained by summing the measured $m_T$ spectra and using the fitted
exponential function (Eq.~\ref{mt}) to extrapolate to full $m_T$. 
For most bins
the necessary correction is small ($\cong 5\%$). The values
of $dn/dy$ at midrapidity ($|y| < 0.6$) are given in Table II.  An
increase of rapidity density with energy is seen for all particles.
The rapidity spectra were parameterised by the sum of two  Gauss
distributions placed symmetrically with respect to midrapidity:
\begin{equation}
\frac {dn} {dy} =
N \cdot \left[\exp  \left( - \frac{(y-y_0)^2}{2 \sigma^2}\right) 
+ \exp  \left( - \frac{(y+y_0)^2}{2 \sigma^2}\right)\right].
\end{equation}
The results of the fits are indicated by the full lines in
Fig. \ref{dndy} and the obtained values of the parameters 
$N, y_0$ and $\sigma$ are given in Table III.
Since both $y_0$ and $\sigma$ increase with
increasing energy, the width of the observed rapidity distributions
clearly increases with energy.
The mean multiplicities in full phase  space were
calculated by integrating  the parametrised rapidity spectra.
The resulting numbers are given in Table II.
The mean multiplicity of $\pi^+$ mesons  given in Table II
was calculated by scaling $\langle \pi^- \rangle$
by the $\pi^+/\pi^-$ ratio 
(0.91, 0.94 and 0.97 at 40, 80 and 158 $A$GeV, respectively)
measured in the  region where both
$dE/dx$ and $tof$ measurements are available.
Similar ratios (within 2\%)  are predicted by the 
Venus model \cite{venus}.
We have also checked, with the Venus model, that the $\pi^+/\pi^-$ ratio
of total multiplicities is, within 1.5\%, equal to the ratio
in the $tof+dE/dx$
acceptance.
Results on $dn/dy$ and inverse slope parameters near midrapidity
at 158 $A$GeV are compatible
with previously published
measurements \cite{na44}.

\section{Energy Dependence}

\begin{figure}[p]
\epsfig{file=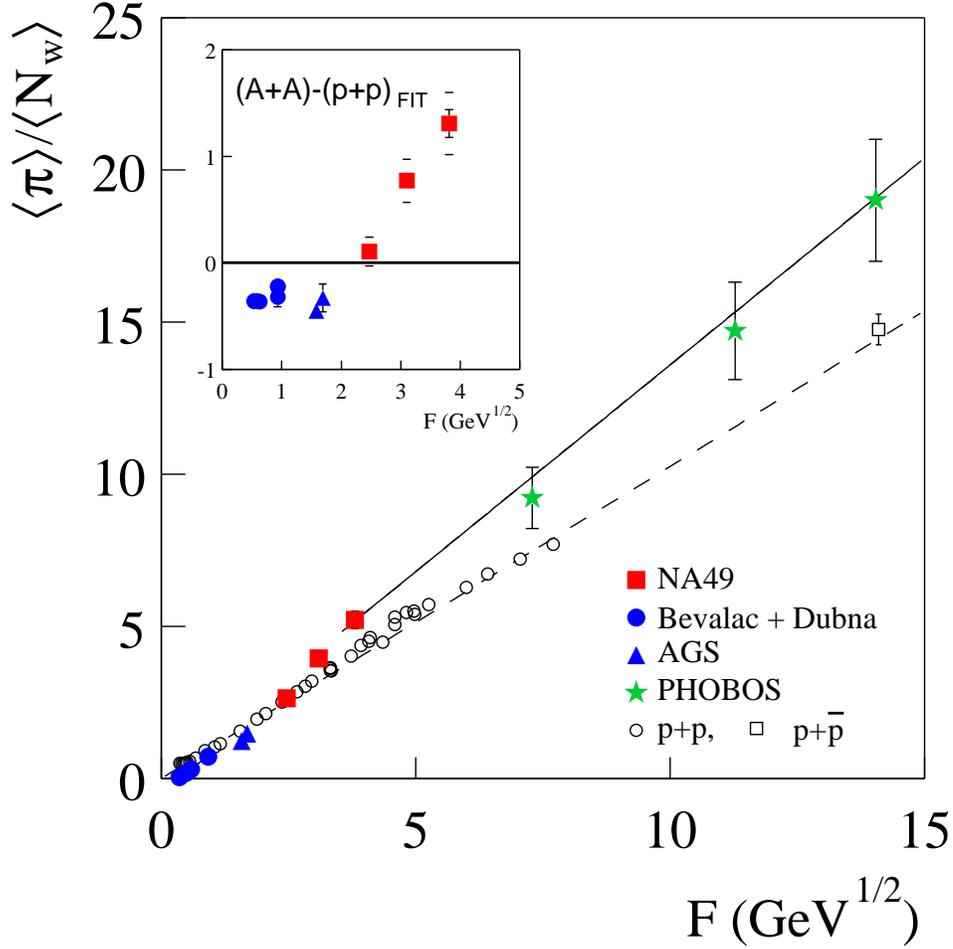,width=14cm}

\vspace{0.5cm}
\caption{Dependence of the total pion multiplicity
per wounded nucleon on Fermi's energy measure $F$ (see text)
for central A+A collisions (closed symbols) and inelastic
$p+p$($\overline{p}$)  interactions 
(open symbols). 
The results of NA49 are indicated by squares.
The full line shows a linear  fit through the origin
to the A+A data at and above
158 $A$GeV.
The insert shows the difference between the results for A+A
collisions and the  straight line 
parametrisation of $p+p$($\overline{p}$) data
(dashed line).
The inner error bars on the NA49 points indicate the statistical
uncertainty and the outer error bars the statistical and
systematic uncertainty added in quadrature.
}
\label{energy}
\end{figure}

The energy dependence of the mean pion multiplicity $\langle \pi
\rangle = 1.5 \cdot (\langle \pi^+ \rangle + \langle \pi^- \rangle)$
is shown in Fig.~\ref{energy}.  In this figure the ratio $\langle \pi
\rangle/\langle N_W \rangle$ is plotted as a function of the collision
energy, expressed by Fermi's measure \cite{Fe:50}: $F \equiv (\sqrt{s_{NN}}
- 2 m_N)^{3/4}/\sqrt{s_{NN}}^{1/4}$, where $\sqrt{s_{NN}}$ is the c.m.s. energy
per nucleon--nucleon pair and $m_N$ the rest mass of the nucleon.
Measurements by NA49 are compared to results from other experiments on
central nucleus--nucleus collisions \cite{GaRo,ags,rhic_pions} and to
a compilation of data from $p+p$($\overline{p}$) interactions (see
references in \cite{GaRo}).

One observes that the mean pion multiplicity in 
$p+p$($\overline{p}$)
interactions is approximately proportional to $F$;
the dashed line in Fig.~\ref{energy} indicates a fit of 
the form $\langle \pi \rangle/\langle N_W \rangle = a \cdot F $ 
to the data, yielding $a = 1.025 \pm 0.005$ GeV$^{-1/2}$. 
For central A+A collisions the dependence is  more
complicated and it can not be fitted by a single linear function
($\chi^2/dof \approx 18$).
Below 40 $A$GeV the ratio $\langle \pi \rangle/\langle N_W \rangle$ in
A+A collisions is lower than in $p+p$ interactions (pion suppression),
while at higher energies $\langle \pi \rangle/\langle N_W \rangle$ is
larger in A+A collisions than in $p+p$($\overline{p}$) interactions
(pion enhancement). In the region between AGS and the lowest SPS
energy (15--40 $A$GeV) the slope changes from $a = 1.01 \pm 0.04$
GeV$^{-1/2}$ ($\chi^2/dof \approx 0.9$) for the fit to the points
up to top AGS energy to $ a = 1.36 \pm 0.03 $ GeV$^{-1/2}$ ($\chi^2/dof
\approx 0.2$) for the fit to the top SPS energy and the RHIC data
points \cite{rhic_pions}. 
The fit to the top SPS and RHIC points is indicated by
the full line in Fig.~\ref{energy}.
The transition from pion suppression to pion enhancement
is demonstrated more clearly in the insert of Fig. \ref{energy}, where
the difference between $\langle \pi \rangle/\langle N_W \rangle$
for A+A collisions and the straight line parametrisation of the $p+p$ data
is plotted as a function of $F$
up to the highest SPS energy.

\begin{figure}[p]
\begin{center}
\epsfig{file=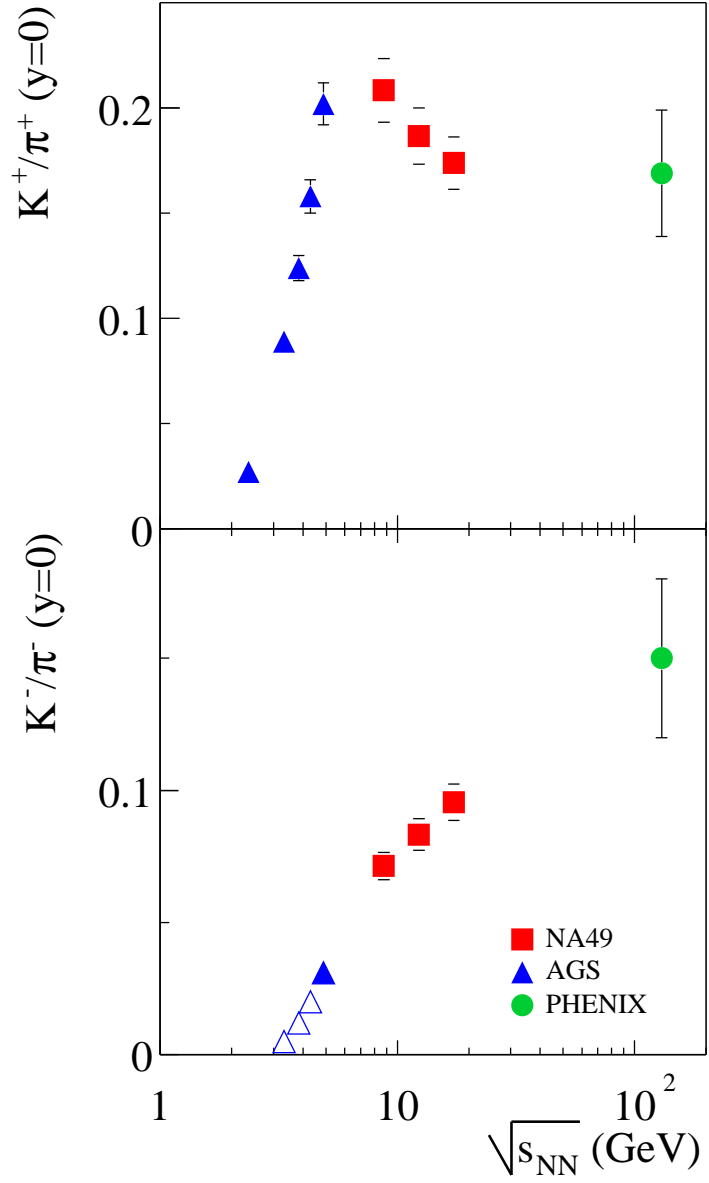,width=10cm}
\end{center}

\vspace{0.5cm}
\caption{Energy dependence of the midrapidity 
$K^+/\pi^+$
and $K^-/\pi^-$
ratios in central Pb+Pb and Au+Au  collisions.
The results of NA49 are indicated by squares.
Open triangles indicate
the A+A results for which preliminary data were used
\protect\cite{Kl:01}.
The errors on the NA49 points are the statistical
and systematic errors added in quadrature. 
The statistical errors are smaller than the symbol size.
}
\label{energy_mid}
\end{figure}

\begin{figure}[p]
\begin{center}
\epsfig{file=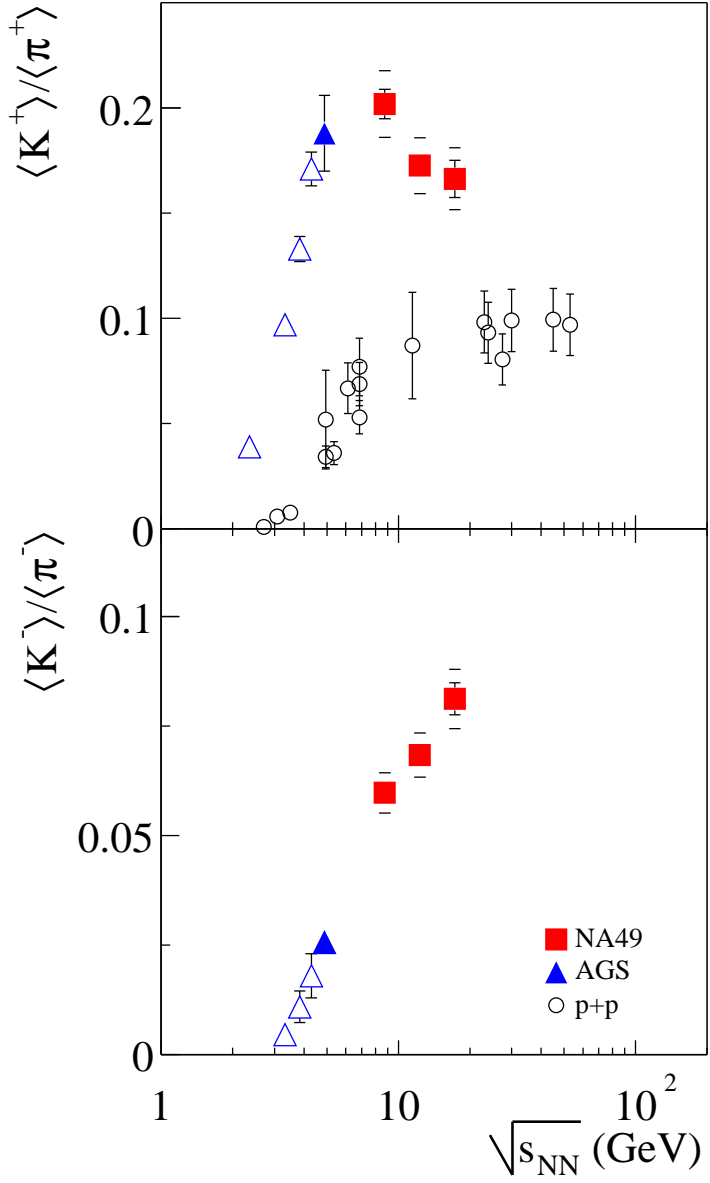,width=10cm}
\end{center}

\vspace{0.5cm}
\caption{
Energy dependence of full phase space
$\langle K^+ \rangle/\langle \pi^+ \rangle$ and  
$\langle K^- \rangle/\langle \pi^- \rangle$
ratios in central Pb+Pb (Au+Au) collisions.
The results of NA49 are indicated by squares.
The data for $p+p$ interactions are shown by open circles
for  comparison.
Open triangles indicate
the A+A results for which a substantial extrapolation was
necessary \protect\cite{extra}. 
The inner error bars on the NA49 points indicate the statistical
uncertainty and the outer error bars the statistical and
systematic uncertainty added in quadrature.
}
\label{energy_4pi}
\end{figure}

Midrapidity  and full phase space kaon to pion ratios are
shown as a function of $\sqrt{s_{NN}}$ in Figs. \ref{energy_mid} and
\ref{energy_4pi} \cite{GaRo,ags,rhic_kaons}, respectively.  A
monotonic increase with $\sqrt{s_{NN}}$ of the $K^-/\pi^-$ ratio is measured.  For 
the $K^+/\pi^+$ ratio, a very different behaviour is observed: a steep
increase in the low (AGS \cite{ags}) energy region is followed by a
maximum around 40 $A$GeV.
The measurement at RHIC indicates that the $K^+/\pi^+$ ratio
stays nearly constant starting from the top SPS energy.
For comparison the results on the $\langle K^+ \rangle/\langle \pi^+
\rangle$ ratio in $p+p$ interactions \cite{GaRo} are also shown in Fig.
\ref{energy_4pi}.  The $p+p$ data have rather large experimental 
uncertainties
\cite{GaRo}, but suggest a monotonic increase of the ratio.  
It should
be noted that the $\langle K^+
\rangle/\langle \pi^+ \rangle$ ratio is expected to be similar
(within about 10\%) for $p+p$, $n+p$ and $n+n$ interactions
\cite{hansen}.

\begin{figure}[p]
\epsfig{file=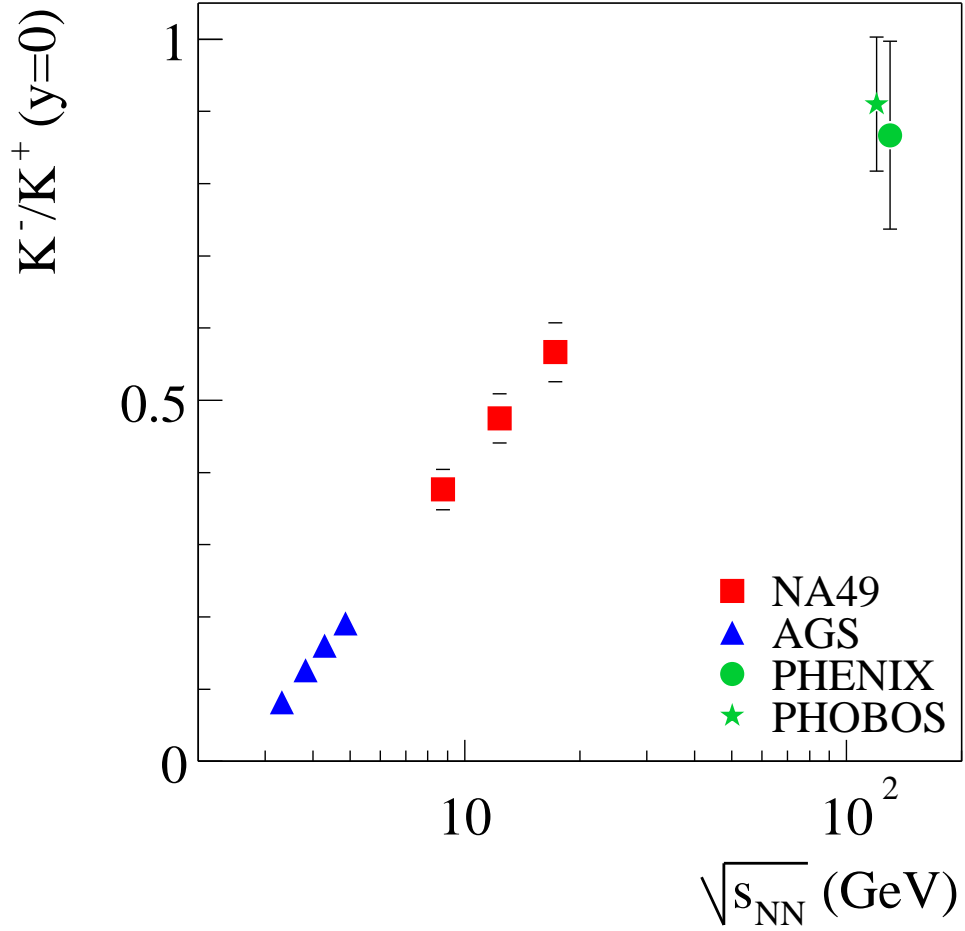,width=14cm}

\vspace{0.5cm}
\caption{Energy dependence of the midrapidity  
 $K^-/K^+$ ratio
in central Pb+Pb (Au+Au) collisions.
The results of NA49 are indicated by squares.
The errors on the NA49 points are the statistical
and systematic errors added in quadrature. 
The statistical errors are smaller than the symbol size.
}
\label{energy_ratio}
\end{figure}

The energy dependence of the $K^-/K^+$ ratio at midrapidity
is shown in Fig. \ref{energy_ratio}.
The ratio increases with $\sqrt{s_{NN}}$ from about 0.15
at low AGS energy \cite{ags} to about 0.5  
at SPS energies and reaches about 0.9
at RHIC \cite{rhic_kaons}.

The difference between the dependence of the $K^+$ and $K^-$ yields
on $\sqrt{s_{NN}}$
can be  attributed  to their different sensitivity
to the baryon density.
Kaons ($K^+$ and $K^0$) carry a dominant fraction 
of all  produced 
$\overline{s}$--quarks
(more than 95\% in Pb+Pb collisions at 158 $A$GeV if open
strangeness is considered).
Therefore the $K^+$ yield ($\langle K^+ \rangle \cong \langle K^0 \rangle$
in approximately isospin symmetric collisions of heavy nuclei)
is nearly proportional to the total strangeness production and only weakly
sensitive to the baryon density.
As a significant fraction of $s$--quarks
(e.g. about 50\% in central Pb+Pb collisions at 158 $A$GeV)
is carried by hyperons, the number of produced antikaons ($K^-$ and
$\overline{K}^0$) is sensitive to both the strangeness yield and  the baryon density.

\begin{figure}[p]
\epsfig{file=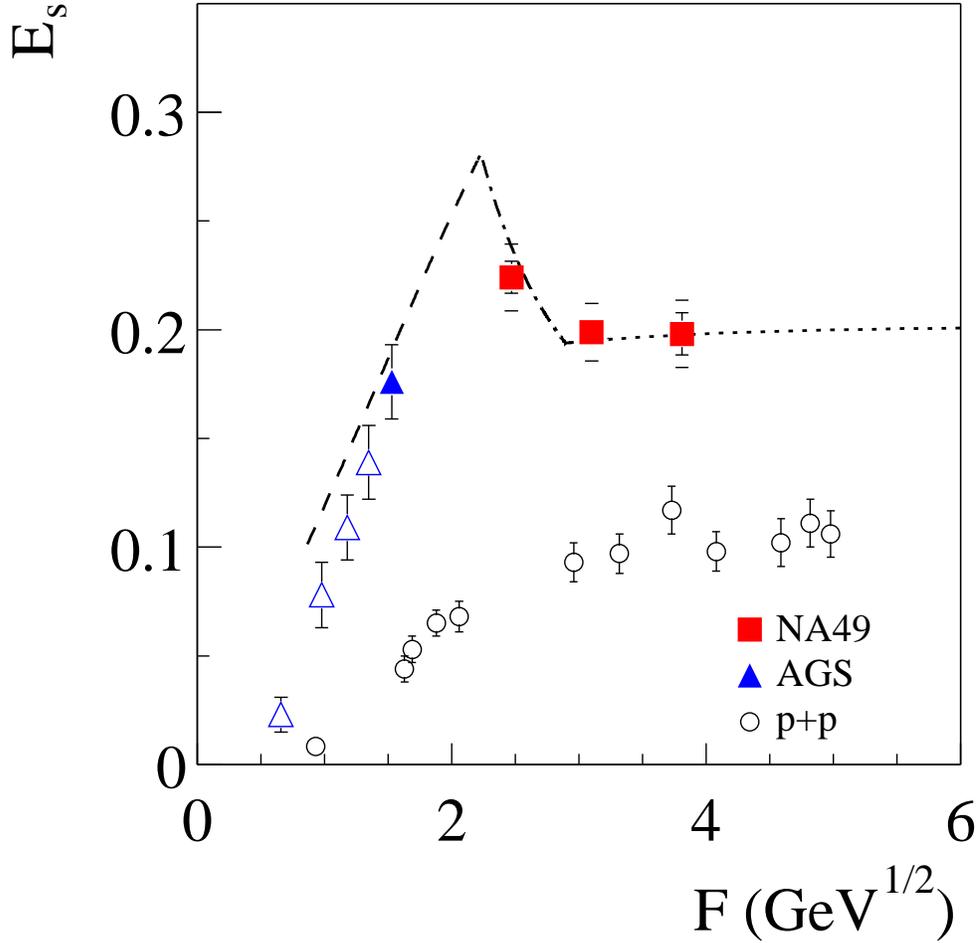,width=14cm}

\vspace{0.5cm}
\caption{Energy dependence of the 
$E_S = (\langle \Lambda \rangle + \langle K + \overline{K} \rangle)/
\langle \pi \rangle$ ratio
in central Pb+Pb (Au+Au) collisions and $p+p$
interactions.
The results of NA49 are indicated by squares.
Open triangles indicate
the A+A results for which a substantial extrapolation was
necessary \protect\cite{extra}.
The experimental results on A+A collisions
 are compared with the predictions
of the Statistical Model of the Early Stage (line)
\protect\cite{GaGo}. 
Different line styles indicate predictions in the energy
domains in which confined matter (dashed line),
mixed phase (dash--dotted line) and QGP (dotted line) are
created at the early stage of the collisions.
The inner error bars on the NA49 points indicate the statistical
uncertainty and the outer error bars the statistical and
systematic uncertainty added in quadrature.
}
\label{es}
\end{figure}

In Fig. \ref{es} an alternative 
measure of the
strangeness to pion ratio,
$E_S = (\langle \Lambda \rangle + \langle K+\overline{K} \rangle)/
\langle \pi \rangle$,
 is plotted
as a function of $F$ for A+A collisions~\cite{ags}
and $p+p$  interactions~\cite{GaRo}.
For A+A collisions the $\Lambda$ multiplicity, when not published,
(e.g. for the NA49 points) was estimated 
as $\langle \Lambda \rangle =
(\langle K^+ \rangle - \langle K^- \rangle$)/0.8,
based on strangeness conservation and  approximate
isospin symmetry of the colliding nuclei \cite{hansen}.   
The wealth of data on $\Lambda$ and $K^0_S$ 
($K^0_S = 0.5\langle K^0 + \overline{K}^0 \rangle \approx 0.5 \langle
K^+ + K^- \rangle$)
production in 
$p+p$ interactions \cite{GaRo} allows a much more precise determination
of $E_S$ (open circles in Fig.~\ref{es})
than
of the $\langle K^+ \rangle/\langle \pi^+ \rangle$ ratio
(open circles in Fig.~\ref{energy_4pi}).
By construction, $E_S$  should be
almost independent (an expected variation of several per cent)
of the charge composition of colliding nuclei.
One may conclude from Figs.~\ref{energy_4pi} and  \ref{es} that
a non--monotonic energy dependence (or a sharp turnover)
of the total strangeness to pion ratio appears to be  a
special property of heavy ion collisions, which is not observed
in elementary interactions.

\section{Comparison with Models}

The energy  dependence of pion and strangeness yields
was discussed within various approaches to nucleus--nucleus
collisions. 
In this section we compare our results with published 
model predictions.  

It was suggested \cite{GaRo,Ga:95} that a transition to 
a deconfined state of matter may cause anomalies
in the energy dependence of pion and strangeness production.
This led to the formulation of 
the Statistical Model of the Early Stage (SMES) \cite{Ga:95,GaGo}
which 
is based on the assumption that 
the system created at the early stage
(be it confined matter or a QGP) is in
equilibrium and a transition from a reaction with purely 
confined matter to a reaction with 
a QGP at the early stage occurs when the transition
temperature $T_C$ is reached.
For $T_C$ values of 170--200 MeV the transition region
ranges between 15--60 $A$GeV \cite{GaGo}.

Due to the  assumed  generalised Fermi--Landau initial
conditions \cite{GaGo,Fe:50,FeLa}
the $\langle \pi \rangle/\langle N_W \rangle$
ratio  
(a measure of entropy per baryon)
increases approximately linear with $F$
outside the transition region.
The slope parameter is proportional to
$g^{1/4}$ \cite{Ga:95}, where $g$ is an effective
number of internal degrees of freedom  at the early stage.
In the transition region a steepening of the
pion energy dependence is expected, because of activation of a large
number of partonic degrees of freedom.
This is, in fact, observed in the data on
central Pb+Pb (Au+Au) collisions, where the steepening starts
in the range 
15--40~$A$GeV (see Fig.~\ref{energy}).
The linear dependence on $F$ is obeyed by the data
at lower and higher energies (including RHIC).
An increase of the slope by a factor of about 1.3 
is measured (see Sect. V), 
which corresponds to an increase of the effective number of
internal degrees of freedom by a factor of 1.3$^4$ $\cong$ 3, 
within the SMES \cite{Ga:95}.

In the SMES model
the $\langle K^+\rangle / \langle \pi^+\rangle$
and $E_S$ ratios are roughly proportional to the total strangeness to
entropy ratio which is assumed to be preserved
from the early stage till freeze--out.  
At low collision energies the
strangeness to entropy ratio steeply increases with collision energy,
due to the low temperature at the early stage ($T < T_C$)
and the high mass of the carriers of strangeness ($m_S \cong 
500 $ MeV, the kaon mass) in the confined state.  
When the transition to a
QGP is crossed ($T > T_C$), the mass of the strangeness carriers is
significantly reduced ($m_S \cong 170$ MeV, the strange quark mass).
Due to the low mass ($m_S < T$) the strangeness yield becomes (approximately)
proportional to the entropy, and the strangeness
to entropy (or pion) ratio is independent of energy.
This leads to a "jump" in the energy
dependence from the larger value for confined matter at $T_C$ to the QGP
value. Thus, within the SMES, the measured non--monotonic energy
dependence of the strangeness to entropy ratio is followed by a
saturation at the QGP value (see Figs. \ref{energy_4pi} and \ref{es})
which is a direct consequence of the onset of deconfinement taking place at
about 40~$A$GeV.

\begin{figure}[p]
\epsfig{file=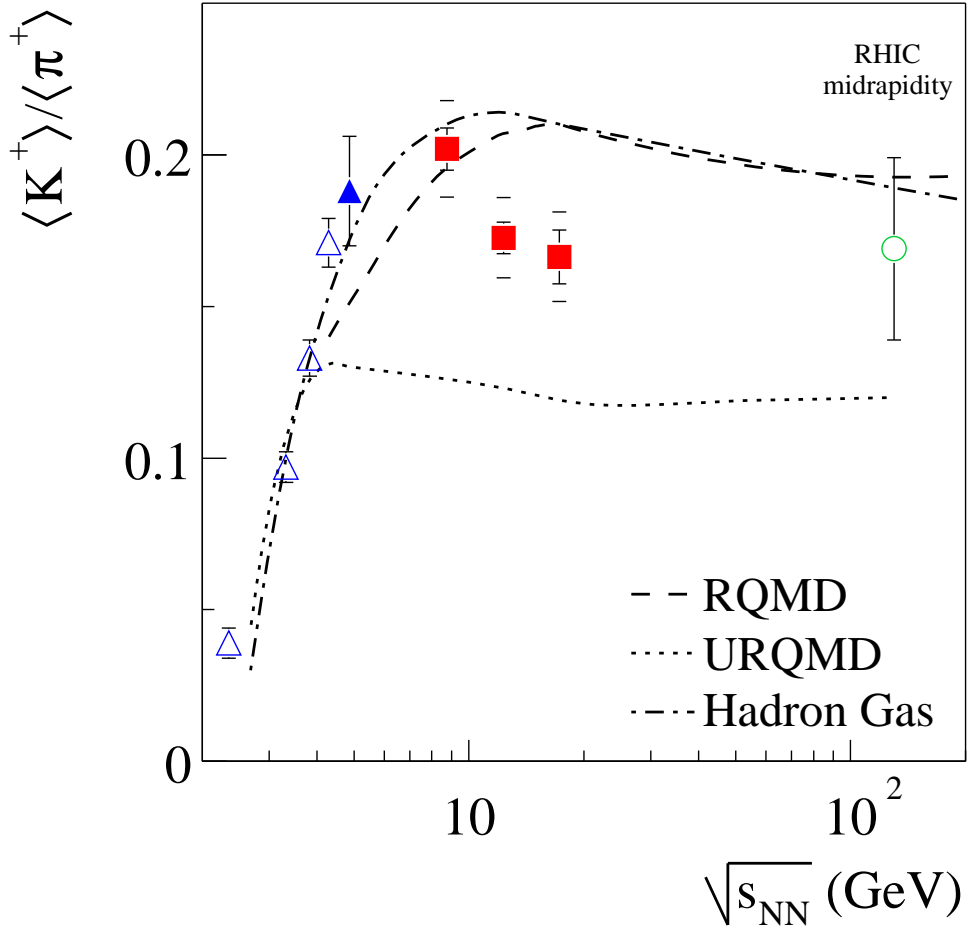,width=14cm}

\vspace{0.5cm}
\caption{Energy dependence of the full phase space 
$\langle K^+ \rangle/\langle \pi^+ \rangle$
ratio in central Pb+Pb and Au+Au collisions.
The experimental results taken from Fig.~\ref{energy_4pi}
are compared to model predictions:
RQMD \protect\cite{RQMD} (dashed line), 
UrQMD \protect\cite{URQMD} (dotted line) and
the Extended Hadron Gas Model \protect\cite{Cl:01} (dash--dotted line).
}
\label{rqmd}
\end{figure}

Numerous models have been developed to explain 
hadron production in reactions of heavy
nuclei without explicitly invoking a transient QGP phase.  The
simplest one is the statistical hadron gas model \cite{Ha:94} where
independent of the collision energy the hadrochemical freeze--out
creates a hadron gas in equilibrium \cite{stock}.  The temperature,
baryon chemical potential and hadronization volume are free parameters
of the model, which are fitted to the data at each energy.  In this
formulation, the hadron gas model does not predict the energy
dependence of hadron production. Recently, an extension of the model
was proposed, in which it is assumed that the values of the thermal
parameters (temperature and baryon chemical potential) evolve smoothly
with the collision energy \cite{Cl:01}.  The energy dependence
calculated within this extended hadron gas model for the $\langle K^+
\rangle/\langle \pi^+ \rangle$ ratio is compared to the experimental
results in Fig. \ref{rqmd}. Due to its construction, the prevailing
trend in the data is reproduced by the model, but the decrease of the
ratio between 40 and 158 $A$GeV is not well described.  The
measured strangeness to pion yield in central Pb+Pb collisions at 158
$A$GeV is about 25\% lower 
than the expectation for the fully equilibrated
hadron gas \cite{Cl:01,Be:98}.

Several dynamical hadron--string models have been developed to study hadron
production in A+A collisions. These models treat the
elementary collisions with a string--hadronic framework as 
a starting point. 
The models are then extended with effects which are
expected to be relevant in A+A collisions (such as string--string
interactions and hadronic rescattering).  
The predictions of the 
RQMD \cite{RQMD,RQMD1} and the UrQMD
\cite{URQMD,URQMD1} models are shown in Fig.~\ref{rqmd}.  
It is seen that
RQMD, like the hadron gas model, fails to describe the decrease of the
$\langle K^+ \rangle/\langle \pi^+ \rangle$ ratio in the SPS energy
range. 
The UrQMD model predicts a ratio which, above $\sqrt{s_{NN}} \cong
5$ GeV, does not show any sizable energy dependence and which is
significantly lower (e.g. by about 40\% at 40 $A$GeV) than the
data. 
This is mainly due to the fact that UrQMD overestimates pion
production at SPS energies by more than 30\% \cite{URQMD1}.

\begin{figure}[p]
\epsfig{file= 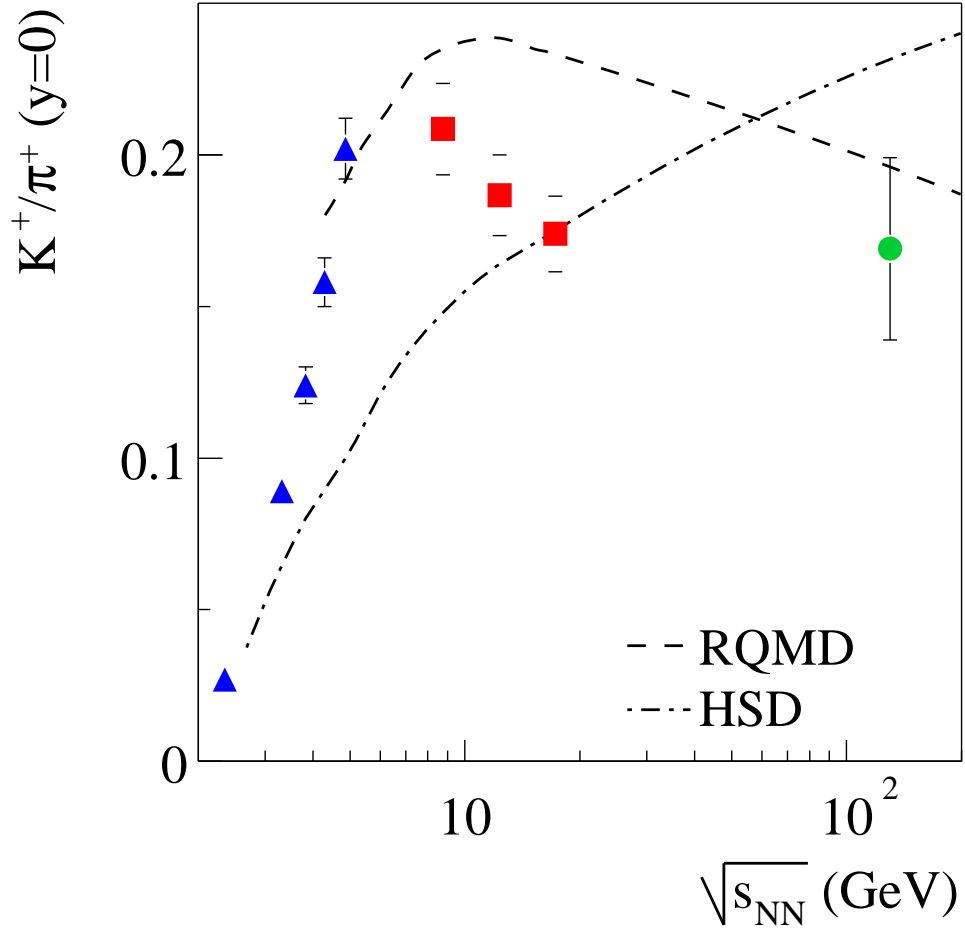,width=14cm}

\vspace{0.5cm}
\caption{Energy dependence of the midrapidity 
$K^+/\pi^+$
ratio in central Pb+Pb and Au+Au  collisions.
The experimental results taken from Fig.~\ref{energy_mid}
are compared with model predictions:
RQMD \protect\cite{RQMD} (dashed line) and
HSD \protect\cite{HSD} (dash--dotted line).
}
\label{hsd}
\end{figure}

The RQMD prediction of the energy dependence
of the $K^+/\pi^+$ ratio at midrapidity is shown in 
Fig~\ref{hsd}. The model also fails to reproduce the experimental data 
both in shape and magnitude.
In addition, Fig.~\ref{hsd}
presents the prediction of the HSD model \cite{HSD}
(another version of the dynamical hadron--string approach)
which shows a monotonic increase of the $K^+/\pi^+$ ratio
with energy. This trend is very different from the measured one.

\section{Summary}
Results on charged pion and kaon production
in central Pb+Pb collisions at 40, 80  and 158 $A$GeV 
are presented.
These are compared with  data at lower and higher
energies as well as with results from $p+p$ interactions.
The mean pion multiplicity per wounded nucleon increases approximately
linearly with $s_{NN}^{1/4}$ with a change of slope starting
in the region 15--40 $A$GeV.
The change from pion suppression  
with respect to $p+p$ interactions, 
as observed at low collision energies, to pion enhancement
at high energies occurs at about 40 $A$GeV.
A non--monotonic energy dependence of the 
ratio of $K^+$ to $\pi^+$ yields is observed, with 
a maximum close to 40 $A$GeV and an indication of a nearly
constant value at higher energies. 
This characteristic energy dependence is not observed in
elementary
interactions and seems to be an unique feature of heavy ion 
collisions.
The measured dependences can be related to an increase of the entropy
production and a decrease of the strangeness to entropy ratio
in central Pb+Pb collisions in the low SPS energy range. 
They can be understood
within the  Statistical Model of the Early Stage of
nucleus--nucleus collisions which assumes
that a transient state of deconfined 
matter is created in Pb+Pb collisions for energies
larger than about 40 $A$GeV.
Currently available
models without this assumption do not 
reproduce the measured energy dependence of pion and
strangeness production equally well.

\newpage
\vspace{1cm}
\noindent
{\bf Acknowledgements}

This work was supported by the Director, Office of Energy Research, 
Division of Nuclear Physics of the Office of High Energy and Nuclear Physics 
of the US Department of Energy (DE-ACO3-76SFOOO98 and DE-FG02-91ER40609), 
the US National Science Foundation, 
the Bundesministerium fur Bildung und Forschung, Germany, 
the Alexander von Humboldt Foundation, 
the UK Engineering and Physical Sciences Research Council, 
the Polish State Committee for Scientific Research (5 P03B 13820 and 2 P03B 02418), 
the Hungarian Scientific Research Foundation (T14920 and T23790),
Hungarian National Science Foundation, OTKA, 
(F034707),
the EC Marie Curie Foundation,
and the Polish-German Foundation.

\newpage

\noindent
{\bf Table I}

\noindent
Numbers of analysed events,  cross sections of selected
central interactions as percentage of total inelastic cross section 
($\sigma^{INEL} = 7.15$ b) and
mean numbers of wounded nucleons for the selected central Pb+Pb collisions 
at 40, 80 and 158 $A$GeV.
The first error given in the last column is statistical, the
second systematic.

\vspace{1.0cm}

\begin{tabular}{|c|c|c|c|}
\hline
          &         &       &              \cr
~Energy [$A$GeV]~               & 
~Number of events~               & 
~~~~$\sigma^{CENT}/\sigma^{INEL}$ [\%]~~~~  &
~~~~~~~$\langle N_W \rangle$~~~~~~~~    \cr
          &         &       &            \cr
\hline
\hline
   40    &  4$\cdot$ 10$^5$   & 7.2 & 349 $\pm$ 1 $\pm$ 5  \cr
   80    &  3$\cdot$ 10$^5$   & 7.2 & 349 $\pm$ 1 $\pm$ 5  \cr
  158    &  4$\cdot$ 10$^5$   & 5.0 & 362 $\pm$ 1 $\pm$ 5  \cr
\hline
\end{tabular}

\newpage

\noindent
{\bf Table II}

\noindent
Inverse slope parameters $T$ of the transverse mass spectra
fitted in the interval 0.2 GeV $< m_T -m <$ 0.7  GeV
at midrapidity ($|y| < 0.1$ for kaons in the  $tof+dE/dx$ analysis, and
$ 0 < y < 0.2 $ for pions),
rapidity densities $dn/dy$ averaged over
the  interval 
$ |y| < 0.6 $
as well as 
total mean multiplicities of $\pi^-$, $\pi^+$, $K^-$ and
$K^+$ mesons produced in central Pb+Pb collisions at
40, 80 and 158 $A$$\cdot$GeV.
The first error is statistical, the second systematic.
Note that $\langle \pi^+ \rangle$ is not directly measured. 

\vspace{1.0cm}

\begin{tabular}{|c|c|c|c|}
\hline
          &         &      &           \cr
          & 
~~~~40 $A$GeV~~~~~   & 
~~~~80 $A$GeV~~~~~   & 
~~~~158 $A$GeV~~~~~     \cr
          &         &       &            \cr
\hline
\hline
  $T(\pi^-)$(MeV) & 169 $\pm$ 2 $\pm$ 10 & 179 $\pm$ 3 $\pm$ 10  
             & 180 $\pm$ 3 $\pm$ 10    \cr
  $T(K^+)$(MeV) & 232 $\pm$ 3 $\pm$ 6   & 230 $\pm$ 5 $\pm$ 6 
 & 232 $\pm$ 4 $\pm$ 6  \cr
  $T(K^-)$(MeV) & 226 $\pm$ 3 $\pm$ 6   & 217 $\pm$ 3 $\pm$ 6 
 & 226 $\pm$ 9 $\pm$ 6   \cr
\hline
~~$dn/dy(\pi^-)$~~ &  106.1 $\pm$ 0.4 $\pm$ 6 & 140.4 $\pm$0.5 $\pm$ 7  & 175.4 $\pm$ 0.7 $\pm$ 9   \cr
~~$dn/dy(\pi^+)$~~ &  96.6 $\pm$ 0.4 $\pm$ 6 & 132.0 $\pm$0.5 $\pm$ 7  & 170.1 $\pm$ 0.7 $\pm$ 9   \cr
~~$dn/dy(K^+)$~~  &  20.1 $\pm$ 0.3 $\pm$  1.0 & 24.6 $\pm$ 0.2 $\pm$ 1.2
 & 29.6 $\pm$ 0.3 $\pm$ 1.5\cr
~~$dn/dy(K^-)$~~  & ~~ 7.58 $\pm$ 0.12 $\pm$ 0.4~~ & 11.7 $\pm$ 0.10 $\pm$ 0.6
& 16.8 $\pm$ 0.2 $\pm$ 0.8 \cr
\hline
  $\langle \pi^- \rangle$ & 322 $\pm$ 3 $\pm$ 16  &  474 $\pm$ 5 $\pm$ 23 
& 639 $\pm$ 17 $\pm$ 31  \cr
  $\langle \pi^+ \rangle$ & 293 $\pm$ 3 $\pm$ 15  &  446 $\pm$ 5 $\pm$ 22 
& 619 $\pm$ 17 $\pm$ 31  \cr
  $\langle K^+ \rangle$ & 59.1 $\pm$ 1.9 $\pm$ 3  & 76.9 $\pm$ 2 $\pm$ 4 
& ~~103.0 $\pm$ 5 $\pm$ 5  ~~\cr
  $\langle K^- \rangle$ & 19.2 $\pm$ 0.5 $\pm$ 1.0  &~~ 32.4 $\pm$ 0.6 $\pm$ 1.6 ~~
& 51.9 $\pm$ 1.9 $\pm$ 3           \cr
\hline
\end{tabular}

\newpage

\noindent
{\bf Table III}

\noindent
Fitted parameters of the two--Gauss parametrization 
(see text, Eq. 2) of
rapidity distributions
measured for 
$\pi^-$,  $K^-$ and
$K^+$ mesons produced in central Pb+Pb collisions at
40, 80 and 158 $A$$\cdot$GeV.
Only statistical errors are given.

\vspace{1.0cm}

\begin{tabular}{|c|c|c|c|}
\hline
          &         &      &           \cr
          & 
~~~~40 $A$GeV~~~~~   & 
~~~~80 $A$GeV~~~~~   & 
~~~~158 $A$GeV~~~~~     \cr
          &         &       &            \cr
\hline
\hline
  $N(\pi^-)$ & 74.0 $\pm$ 0.5  & 97.0  $\pm$ 0.7  & 107.6 $\pm$ 1.8   \cr
  $N(K^+)$ & 16.2 $\pm$ 0.4    & 19.3  $\pm$ 0.3  & 23.4  $\pm$ 0.6   \cr
  $N(K^-)$ & 6.03 $\pm$ 0.13   & 9.16  $\pm$ 0.12 & 12.8  $\pm$ 0.3   \cr
\hline
~~$\sigma(\pi^-)$~~ & 0.872 $\pm$ 0.005 & 0.974 $\pm$ 0.007 
                    & 1.18 $\pm$ 0.02   \cr
~~$\sigma(K^+)$~~  & 0.725 $\pm$ 0.016 & 0.792  $\pm$ 0.018  & 0.88 $\pm$ 0.04 \cr
~~$\sigma(K^-)$~~  & 0.635 $\pm$ 0.011 & 0.705  $\pm$ 0.010  & 0.81 $\pm$ 0.02 \cr
\hline
  $y_0(\pi^-)$ & 0.666 $\pm$ 0.006 & 0.756 $\pm$ 0.006 
               & 0.72 $\pm$ 0.02  \cr
  $y_0(K^+)$ & 0.694 $\pm$ 0.008   & 0.742 $\pm$ 0.008 & 0.839  $\pm$ 0.012  \cr
  $y_0(K^-)$ & 0.569 $\pm$ 0.010   & 0.668 $\pm$ 0.005 & 0.727  $\pm$ 0.010  \cr
\hline
\end{tabular}
\newpage

\newpage

\newpage

\newpage

\newpage

\newpage

\newpage

\newpage

\newpage

\end{document}